\documentclass[aps,prl,amsmath,amssymb,amsfonts,superscriptaddress,floatfix,twocolumn]{revtex4}

\usepackage{graphicx,color}
\usepackage{braket}

\begin{document}

\title{Shadows of Anyons}

\author{J. \surname{Haegeman}}
\affiliation{Department of Physics and Astronomy, University of Ghent, Krijgslaan 281 S9, B-9000 Ghent, Belgium}
\author{V. \surname{Zauner}}
\affiliation{Vienna Center for Quantum Technology, University of Vienna, Boltzmanngasse 5, 1090 Wien, Austria}
\author{N. \surname{Schuch}}
\affiliation{JARA Institute for Quantum Information, RWTH Aachen University, D-52056 Aachen, Germany}
\author{F. \surname{Verstraete}}
\affiliation{Department of Physics and Astronomy, University of Ghent, Krijgslaan 281 S9, B-9000 Ghent, Belgium}
\affiliation{Vienna Center for Quantum Technology, University of Vienna, Boltzmanngasse 5, 1090 Wien, Austria}

\begin{abstract}
The eigenvalue structure of the quantum transfer matrix is known to encode essential information about the elementary excitations. Here we study transfer matrices of quantum states in a topological phase using the tensor network formalism. We demonstrate that topological quantum order requires a particular type of `symmetry breaking' for the fixed point subspace of the transfer matrix, and relate physical anyon excitations to domain wall excitations at the level of the transfer matrix. A topological phase transition to a trivial phase triggers a change in the fixed point subspace to either a larger or smaller symmetry and we explain how this relates to a condensation or confinement of the corresponding anyon sectors. The tensor network formalism enables us to determine the structure of the topological sectors in two-dimensional gapped phases very efficiently, therefore opening novel avenues for studying fundamental open questions related to anyon condensation.
\end{abstract}

\maketitle

\section{Introduction}

One of the major recent advances in the understanding of strongly correlated quantum many body systems has been the investigation of quantum entanglement in terms of area laws \cite{2007JSMTE..08...24H,2010RvMP...82..277E}, the entanglement spectrum \cite{PhysRevLett.101.010504} and the associated entanglement Hamiltonian \cite{PhysRevB.83.245134,PhysRevLett.111.090501}. The structure of entanglement in gapped quantum systems has resulted in the powerful parameterization of quantum ground states in terms of so-called tensor network states, such as matrix product states (MPS) \cite{1992CMaPh.144..443F,1995PhRvL..75.3537O} or their higher dimensional analogues, projected entangled pair states (PEPS) \cite{2004cond.mat..7066V,2008AdPhy..57..143V}. In a translation invariant tensor network state, the entanglement features can be extracted from the leading eigenvector of the so-called (quantum) transfer matrix, which naturally appears whenever the quantum state is mapped to a classical partition function (using \textit{e.g.} a Trotter decomposition) and then sliced along the virtual direction, \textit{i.e.} the Trotter direction or direction of imaginary time \cite{Suzuki2003334}. In a very recent publication \cite{2014arXiv1408.5140Z}, we have observed that the other eigenvalues of the transfer matrix also contain useful information, \textit{i.e.}\ they contain crucial information about the elementary excitations and the corresponding dispersion relations of the system. This is in many ways surprising as this information is completely encoded in the ground state description of the system, a priori without any information about the original Hamiltonian for which it was the exact or approximate ground state.

This relation is extremely useful for the case of two-dimensional systems, where systematic methods to extract the dispersion relation of elementary excitations are virtually non-existent. This was illustrated in Ref.~\onlinecite{2014arXiv1408.5140Z} by studying the Affleck-Kennedy-Lieb-Tasaki model \cite{1987PhRvL..59..799A,1988CMaPh.115..477A}. However, for the case of two-dimensional quantum spin systems exhibiting topological quantum order, the excitation spectrum can be much richer. In contrast to trivial phases, where the elementary excitations can provably be created as Bloch waves of localized perturbations \cite{PhysRevLett.111.080401}, elementary excitations in topological phases are typically anyons that come with ``strings attached'' \footnote{Already for one-dimensional systems, there are well known Hamiltonians such as the Ising model or the Lieb-Liniger model \cite{1963PhRv..130.1605L} whose elementary excitations are topologically nontrivial. In terms of matrix product states, those excitations can be represented with a Bloch like ansatz where an extra half-infinite string operator is attached to the local perturbation \cite{Mandelstam:1975aa,2012PhRvB..85j0408H,PhysRevLett.111.020402,PhysRevD.88.085030}. In the case of symmetry breaking, this string represents a product of local unitary operators which maps one ground state to another one.% As discussed extensively in Ref.~\onlinecite{2014arXiv1408.5140Z}, there is in that case a clear connection between the dispersion relation of these topologically nontrivial excitations and the eigenvalues of a so-called ``mixed'' transfer matrix.
}.

The PEPS representation of topologically ordered ground states, either as variational approximation \cite{PhysRevLett.106.107203,2012NJPh...14b5005S} or as exact description of the Levin-Wen model wave functions \cite{2005PhRvB..71d5110L,2009PhRvB..79h5119B,2009PhRvB..79h5118G}, has been well established. Here we study the full spectrum of the transfer matrix of topologically ordered PEPS \cite{PhysRevLett.111.090501} on the infinite plane or cylinder. Much like the prisoners in Plato's cave, we observe one-dimensional domain walls in the spectrum of the transfer matrix as shadows of the true anyons in the two-dimensional world. By clarifying how the different anyon sectors are manifested at the virtual level, we can probe the dispersion relation of single anyon states. We discuss how the presence of anyons, and thus of topological order, requires a particular type of symmetry breaking in the fixed point subspace of the transfer matrix and how anyon condensation or confinement \cite{PhysRevB.79.045316,1367-2630-14-1-015004} is reflected in these virtual description. We illustrate our results by studying the PEPS description of the toric code model with string tension and the resonating valence bond state. Our results also confirm that we can construct approximate eigenvectors of PEPS transfer matrices using the matrix product ansatz developed for one-dimensional quantum Hamiltonians in \cite{2012PhRvB..85c5130P,2012PhRvB..85j0408H}.

\section{Results}
\subsection{Topological order in PEPS}

\begin{figure}
\includegraphics[width=\columnwidth]{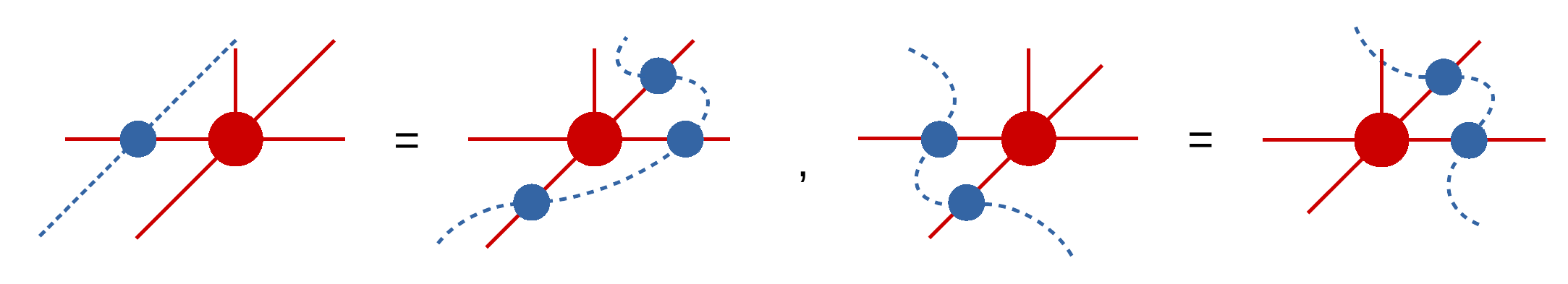}
\caption{\label{fig:pulling}
The pulling through equation between the PEPS tensor (red) and MPO tensor (blue) is characteristic for topological order in PEPS.}
\end{figure}

The convience of the tensor network description of quantum states is that the global, topological properties of the state are reflected in the symmetries of the local tensors. Since topological phases are not characterized by local order parameters, these symmetries act purely on the virtual levels of the tensors. In particular, it was recently established that topological order in PEPS can be characterized by the existence of matrix product operators (MPO) which can be pulled through the lattice at the virtual level (see Fig.~\ref{fig:pulling}) \cite{2010AnPhy.325.2153S,2014AnPhy.351..447B,2014arXiv1409.2150B}. Closed MPO loops around a topologically trivial region define the invariant subspace on which the PEPS tensors are supported and in this way characterizes the topological properties of the state, such as the topological corrections to the entanglement entropy. They act as virtual operators $O_i$ and satisfy a fusion algebra $O_i O_j=\sum_{k} N_{i,j}^{k} O_k$. Indeed, as shown in Ref.~\onlinecite{2014arXiv1409.2150B}, for the case of the Levin-Wen string net models, the different MPOs $O_i$ can be associated to and labeled by the different string types $i=1,\ldots,N$ of the input category that defines the string net model. A PEPS in a trivial phase is characterized by a single MPO $O_1$ that acts as the identity in the relevant subspace. Another interesting case is that of the quantum double models, which can be described using $\mathsf{G}$-injective PEPS \cite{2010AnPhy.325.2153S}. This is a special case of the formalism of Ref.~\onlinecite{2014arXiv1409.2150B} where the MPOs are labeled by the group elements $g\in\mathsf{G}$ and correspond to representations $O_g = U(g)$ of the group action at the virtual level. The pulling through condition is satisfied since the tensors are only supported on the invariant subspace defined by the projector $P=\sum_{g} O_g$.

While the pulling-through conditions ensure that the presence of an MPO string cannot be detected locally, noncontractible MPO loops can have global effects, such as adding a nontrivial flux in the system, and can therefore be used to map one ground state to another one. The relevance of these virtual MPOs is that also away from the the renormalization group (RG) fixed point ---where the physical string operators are spread out and not exactly known \cite{2005PhRvB..72d5141H}--- the MPOs at the virtual level remain strictly local and the `pulling-through' symmetry of the PEPS tensor is exactly preserved.

\subsection{Anyon excitations in the PEPS picture}

Having a translation invariant PEPS description of the ground state of a topological phase, one can easily argue that a suitable ansatz to model single anyon excitations is obtained by modifying the ground state tensors in a local region (e.g. a single site) and attaching a half infinite string to it, which is exactly given by this MPO at the virtual level. The MPO will give rise to the non-trivial braiding statistics of these excitations, while the `pulling-through` assures that the bulk of the string is locally unobservable, so that the energy density is left at its ground state value sufficiently far away from the end point. Away from the RG fixed point, these excitations will disperse and a proper eigenstate can be obtained by building a momentum superposition with momentum $k_x$ and $k_y$ in the $x$ and $y$ direction.

\begin{figure}
\includegraphics[width=0.75\columnwidth]{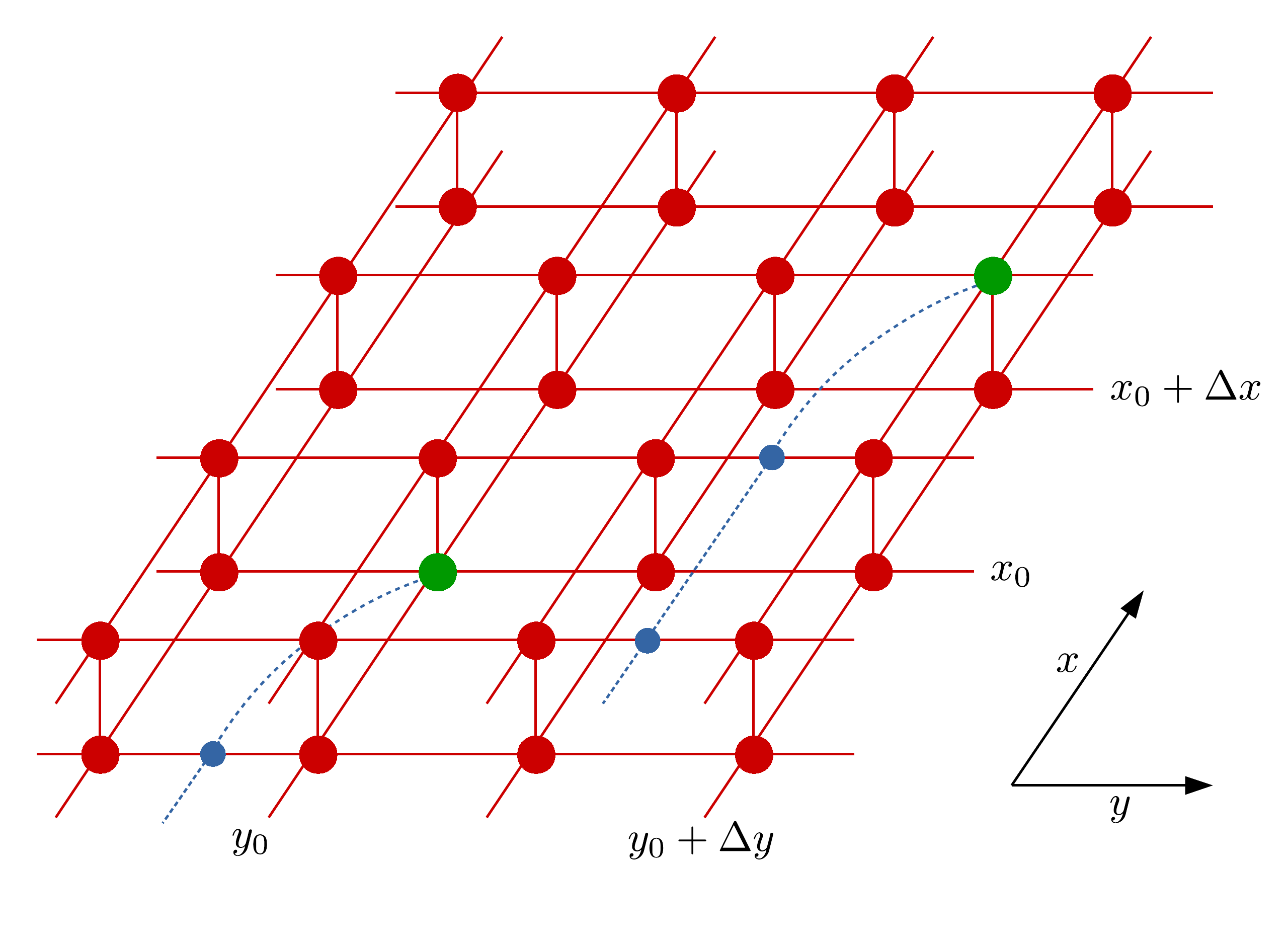}
\caption{\label{fig:peps}
A PEPS ansatz for a topologically non-trivial excitation is constructed from the tensors of the PEPS ground state (red) and a local perturbation (green) with an MPO (blue string) attached. By building momentum superpositions, the variational excitation energy will depend on the overlap  with any difference $\Delta x$ and $\Delta y$ between the string end points in ket and bra, as shown here.}
\end{figure}

Note that the topological quantum numbers of the anyon excitations are not completely specified by the string type $i$, but are determined by structure of the excitation tensor in the ansatz of Fig.~\ref{fig:peps}. For the case of the quantum doubles, it has been shown that the string type corresponds to the magnetic flux, whereas the charge quantum number is determined by the representation space on which the local tensor is supported \cite{2010AnPhy.325.2153S}. A complete characterization of the different anyon sectors in the PEPS formalism would take us to far and is presented elsewhere \footnote{N.~Bultinck, M.~Mari\"{e}n, D.~Williamson, J.~Haegeman, F.~Verstraete, \textit{in preparation}}.

\subsection{Transfer matrix: symmetries and domain walls}

\begin{figure}
\includegraphics[width=1.0\columnwidth]{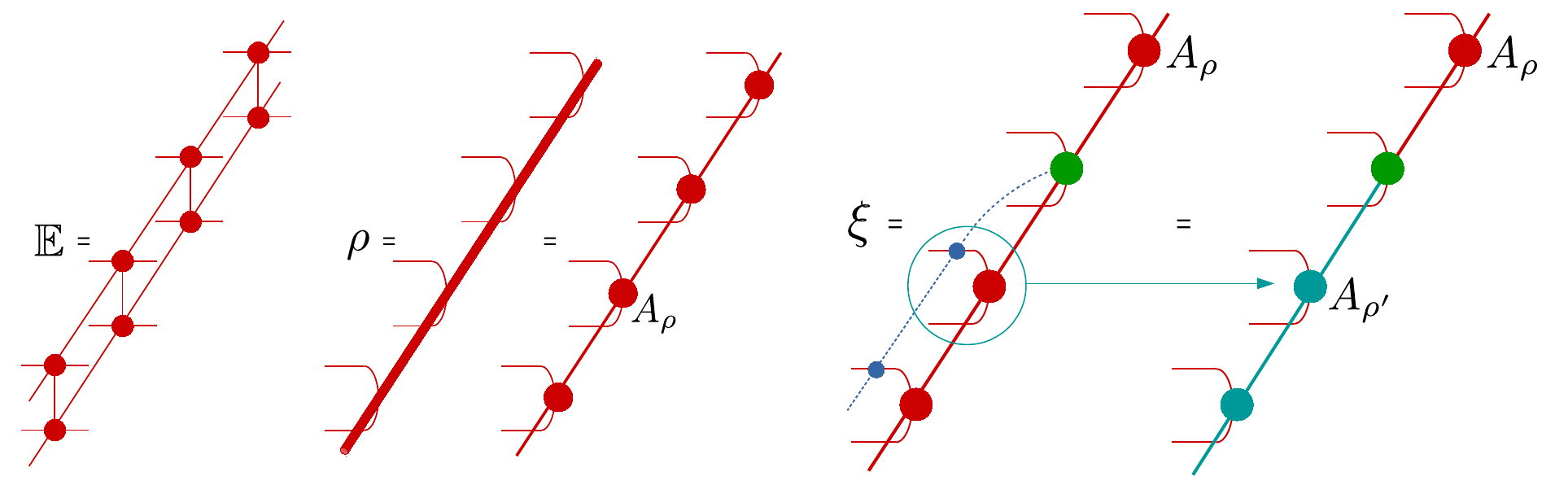}
\caption{\label{fig:tm}
The (normal) transfer matrix $\mathbb{E}$ and one if its fixed points $\rho$, which we represent as an infinite matrix product state. Excited states $\xi$ of the transfer matrix can be constructed by attaching a half infinite MPO to a local excitation tensor (green), which is equivalent to interpolating between the MPS tensors $A_\rho$ and $A_{\rho'}$ of two different fixed points.}
\end{figure}

As in Ref.~\onlinecite{2014arXiv1408.5140Z}, one can now argue that the dominant contribution to the variational dispersion relation is coming from the normalization of these states, which is given by the sum of overlaps of ket and bra with string end points at different positions $(x_0,y_0)$ and $(x_0+\Delta x,y_0+\Delta y)$, as illustrated in Fig.~\ref{fig:peps}. If we orient the strings along the $x$ direction and first contract the tensor network along the $y$ direction, the central object will be the transfer matrix $\mathbb{E}$ in the $x$ direction, as defined in Fig.~\ref{fig:tm}. The pulling through condition of Fig.~\ref{fig:pulling} ensures that $\mathbb{E}$ commutes with infinite MPO strings along the $x$ axis in the ket and bra level separately. We thus obtain $[O_i\otimes\openone,\mathbb{E}]=[\openone\otimes \overline{O}_{i},\mathbb{E}]=0$, $\forall i=1,\ldots,N$ where $O_i$ now denotes an infinite MPO of type $i$ along the $x$ axis. Normalization of the PEPS ground state requires that the largest eigenvalue of $\mathbb{E}$ is $1$, and the infinite power of the transfer matrix $\mathbb{E}$ in the overlap of Fig.~\ref{fig:peps} in the regions $y<y_0$ and $y>y_0+\Delta y$ can be replaced by its left and right fixed point $\sigma$ and $\rho$, which we represent as an infinite MPS with matrices $A_{\sigma}$ and $A_{\rho}$ (Fig.~\ref{fig:tm}).

The transfer matrix $\mathbb{E}$ can have a degenerate fixed point structure, since for a given right fixed point $\rho$, one can build other fixed points $\rho'=O_i\rho O_j^\dagger$ for all $i,j=1,\ldots,N$. One could expect that this generally gives rise to an $N^2$-dimensional fixed point subspace. However, at the RG fixed point of the topological phase, we can easily check that the fixed point subspace of $\mathbb{E}$ is exactly spanned by $\rho_k=O_k$ for $k=0,\ldots,N$, and is thus only $N$-dimensional. The degeneracy and labeling of the fixed point subspace remains intact throughout the topological phase, even though $\rho_k$ will no longer exactly equal $O_k$. This implies, in particular, that $O_i \rho_k O_i^\dagger$ can be expanded into a linear combination of $\sum_{l} c_{kl}\rho_l$ with $c_{k,k}=1$, compatible with the fact that fusing $i \times k \times \overline{\imath}$ will have a fusion channel $k$. We now argue why this property is required in order to support anyonic excitations with a half-infinite virtual string of type $i$.

Contracting the tensor network in Fig.~\ref{fig:peps} from right to left up to position $y_0+\Delta y$ gives rise to some boundary state $\rho$ in the fixed point subspace, whose precise choice is set by the boundary conditions at $y=+\infty$. The topological invariance ensures that this choice has no effect on local expectation values. As we now further contract from right to left, we pass the position $y_0+\Delta y$ containing the excitation in the bra level. Here, the boundary state is perturbed locally at $x=x_0+\Delta x$. In addition, it will be acted upon by a half infinite MPO string of type $i$, which has the effect of changing the MPS tensors from $A_{\rho}$ to $A_{\rho'}$ with $\rho'=\rho O_i^\dagger$ in the region $x<x_0+\Delta x$. At the level of the transfer matrix, the boundary state now takes the form of a domain wall interpolating between the two different fixed points. Because of translation invariance, all overlaps corresponding to the momentum superposition in the $x$ direction can be summed and the resulting state takes the form of a topologically non-trivial state with momentum $k_x$, similar to the domain wall excitation ansatz used for one-dimensional Hamiltonians in Ref.~\onlinecite{2012PhRvB..85j0408H}. We thus have to consider the spectrum of topologically non-trivial eigenstates $\xi_{j}$ (Fig.~\ref{fig:tm}) of $\mathbb{E}$ with momentum $k_x$. Indeed, further contracting up to $y_0$ yields the $k_x$-dependent eigenvalues $\lambda_j(k_x)$ of $\xi_j$ to some power $\Delta y$, which dictates the $k_y$ dependence of the dispersion relation of the corresponding physical excitation.

At point $y_0$, the boundary state is acted upon with a second half-infinite string (also with momentum $k_x$ in the $x$ direction), now in the ket level. After that, the state is collapsed onto the topologically trivial left fixed point $\sigma$. In order to have a non-vanishing overlap, the corresponding boundary state ---with half-infinite strings in both the ket and bra level--- should have a contribution in the trivial sector. This will be true if the property stated above is satisfied, i.e. $O_i \rho O_i^\dagger = \rho + \ldots$ for all $\rho$.

As we perturb the state out of the topological phase, there are two ways in which the existence of topologically non-trivial excitations can break down. If the transfer matrix $\mathbb{E}$ has a unique (maximally `symmetric') fixed point which is invariant under the action of any string $O_i$, then the states $\xi$ interpolate between the same fixed point left and right and are thus  indistinguishable from topologically trivial local perturbations (without string). This scenario is realized when the corresponding anyon has condensed into the ground state \cite{PhysRevB.79.045316,1367-2630-14-1-015004}. A second possibility is that the fixed point structure of $\mathbb{E}$ has an even larger `symmetry breaking', in such a way that the state with half-infinite strings in ket and bra level is still completely topologically non-trivial and has zero overlap with the left fixed point $\sigma$. More specifically, the fidelity per site, defined as the largest eigenvalue $\lambda$ of $\sum_s A^s_{\rho''} \otimes A^s_{\sigma}$ with $\rho''=O_i \rho O_i^\dagger$, satisfies $\lvert\lambda\rvert=\mathrm{e}^{-t}<1$ and the normalization of the state goes down as $\mathrm{e}^{-t L}$ with $L$ the length of the strings. Hence, the nonzero value of $t$ acts as string tension and only bound states of two excitations connected by a finite string can exist, corresponding to the mechanism of confinement.

In the generic case, there can of course be several transitions corresponding to e.g. the condensation of only some anyonic sectors and the induced confinement of other anyon sectors. Note that, since topological phase transitions correspond to symmetry breaking phase transitions at the virtual level of the PEPS description, we can also find virtual order parameters, as illustrated in the toric code example. A more in-depth study of these aspects of anyon condensation within the framework of PEPS will be provided elsewhere \footnote{J. Haegeman, N. Schuch, F. Verstraete, \textit{in preparation}.}.

\subsection{Mixed transfer matrix and momentum fractionalization on the cylinder}
\begin{figure}
\includegraphics[width=\columnwidth]{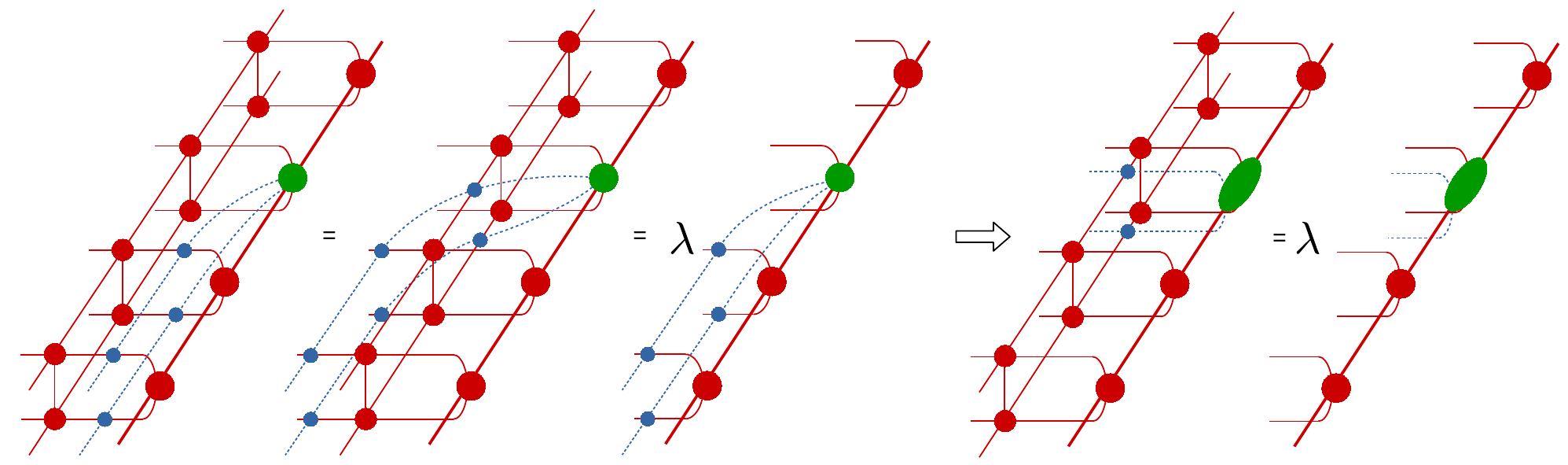}
\caption{\label{fig:alternative}
Rewriting the eigenvalue equation for topologically non-trivial excitations of the normal transfer matrix as a normal eigenvalue equation for a so-called `mixed' transfer matrix having additional MPO indices.}
\end{figure}

In the above discussion, we have explained how information about anyon excitations in topological phases can be obtained from the topologically non-trivial excitations of the translation invariant transfer matrix, which has a degenerate fixed point subspace in the case of topological order. Fig.~\ref{fig:alternative} motivates an alternative approach. By using the pulling through property of the MPO, we can rewrite the eigenvalue equation for a topologically non-trivial excitation of $\mathbb{E}$ as a normal (topologically trivial) eigenvalue problem for a so-called mixed transfer matrix. The latter is threaded by an MPO string and is thus defined on a larger vector space corresponding to the presence of additional MPO indices. Physically, we are effectively rotating the MPO strings attached to the anyon excitation to lie along the $y$ direction.

\begin{figure}
\includegraphics[width=0.9\columnwidth]{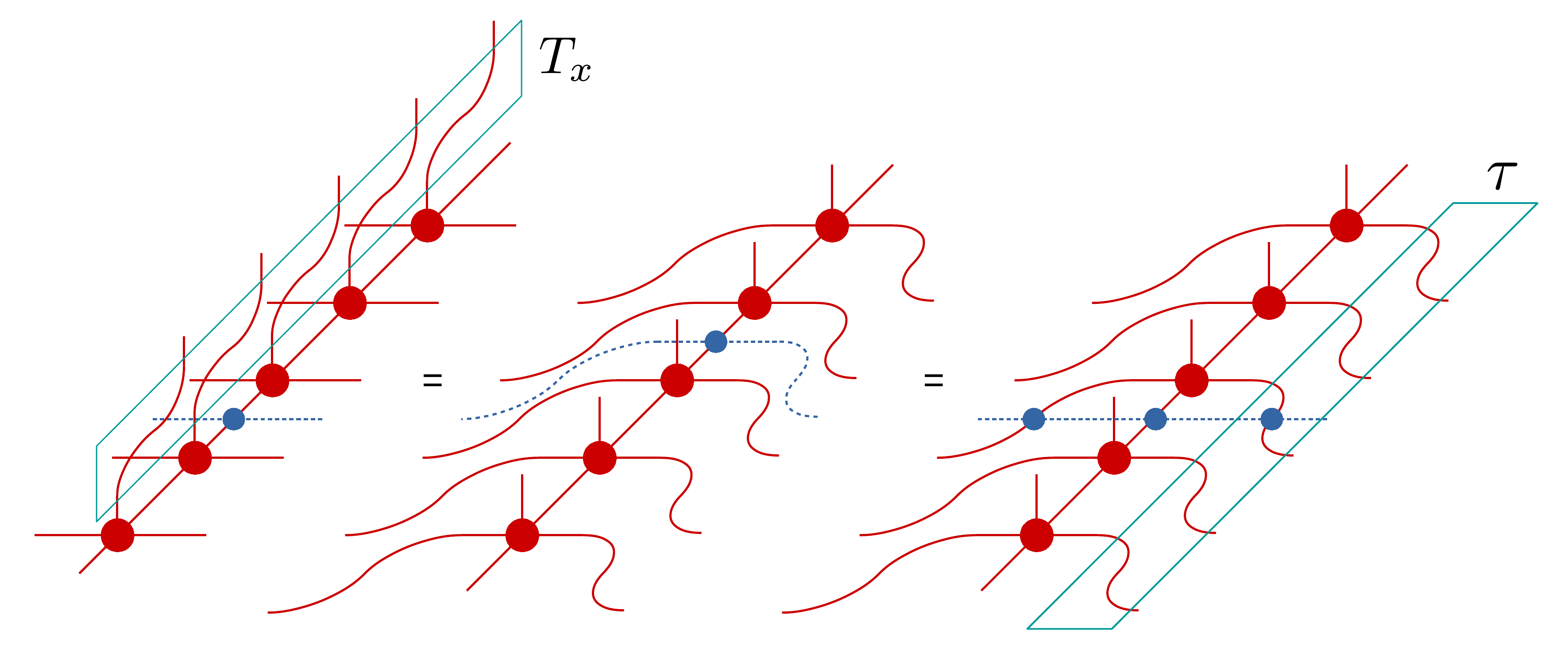}
\caption{\label{fig:virtualtranslation}
The physical translation operator $T_x$ has a non-trivial representation $\tau$ on the virtual level.}
\end{figure}

By doing so, we can make the $x$-direction finite and periodic, which allows to work on a cylinder with finite circumference. The fixed points of these mixed transfer matrices were first studied in Ref.~\onlinecite{PhysRevLett.111.090501}. In this case, physical translations in the $x$ direction have a representation as modified translation operators at the virtual level of the transfer matrix with a non-trivial action on the extra MPO indices [see Fig.~\ref{fig:virtualtranslation}]. This results in a momentum label which can have a fractional discretization in the circumference of the cylinder, similar to what is observed in the case of momentum polarization \cite{2013PhRvB..88s5412T}. We elaborate on this aspect in the Supplementary Material.

\subsection{Anyon excitations and topological phase transitions in the toric code model with string tension}
Let us now illustrate this approach using the toric code ground state, to which we apply a local filtering 
\begin{displaymath}
\prod_e \mathrm{e}^{\frac{\beta_x \sigma_e^x + \beta_z \sigma_e^z}{4}}\prod_{v} (1+\prod_{e \ni v} \sigma^x_{e}) \prod_{p}(1+\prod_{e \in \partial p} \sigma^z_{e}) \ket{\Omega},
\end{displaymath}
with $\ket{\Omega}$ the fully polarized spin state $\ket{\Omega}=\bigotimes_{e} \ket{\uparrow}_{e}$.
The filtering induces dynamics to the elementary excitations of the toric code ground state and can drive the system into a trivial phase. Along the coordination axes $\beta_x=0$ or $\beta_z=0$, it can be interpreted as string tension in either the group or representation basis and can be solved exactly \cite{PhysRevB.76.224421,PhysRevB.77.054433}. The full two-dimensional phase diagram as function of $\beta_x$ and $\beta_z$ was studied in Ref.~\onlinecite{2014arXiv1407.1025H} using a fidelity approach \cite{2008PhRvL.100h0601Z,2008JPhA...41O2001Z,PhysRevA.78.010301}.

The PEPS representation of the toric code ground state was first constructed in Ref.~\onlinecite{2006PhRvL..96v0601V} and its properties were discussed at length in Ref.~\onlinecite{2010AnPhy.325.2153S}. The PEPS tensors satisfy the property of $\mathsf{G}$-injectivity, where in this case $\mathsf{G}=\mathbb{Z}_{2}$. For this particular case, it means that the MPO projector $P$ is given as $O_0+O_1=\openone^{\otimes L} + Z^{\otimes L}$, with $L$ the length of the MPO string and $Z$ a representation of the non-trivial element of $\mathbb{Z}_2$. Correspondingly, the transfer matrix $\mathbb{E}$ along the $x$ direction has the global symmetry $(\openone\otimes\openone,Z^{\otimes N_x} \otimes \openone,\openone\otimes Z^{\otimes N_x},Z^{\otimes N_x} \otimes Z^{\otimes N_x})$ with $N_x$ the number of sites in the $x$-direction. The filtering operation is applied at the physical level of the PEPS and has no effect on any of these properties. It does however influence the manifestation of the symmetry in the fixed point subspace of the transfer matrix.

\begin{figure}
\includegraphics[width=\columnwidth]{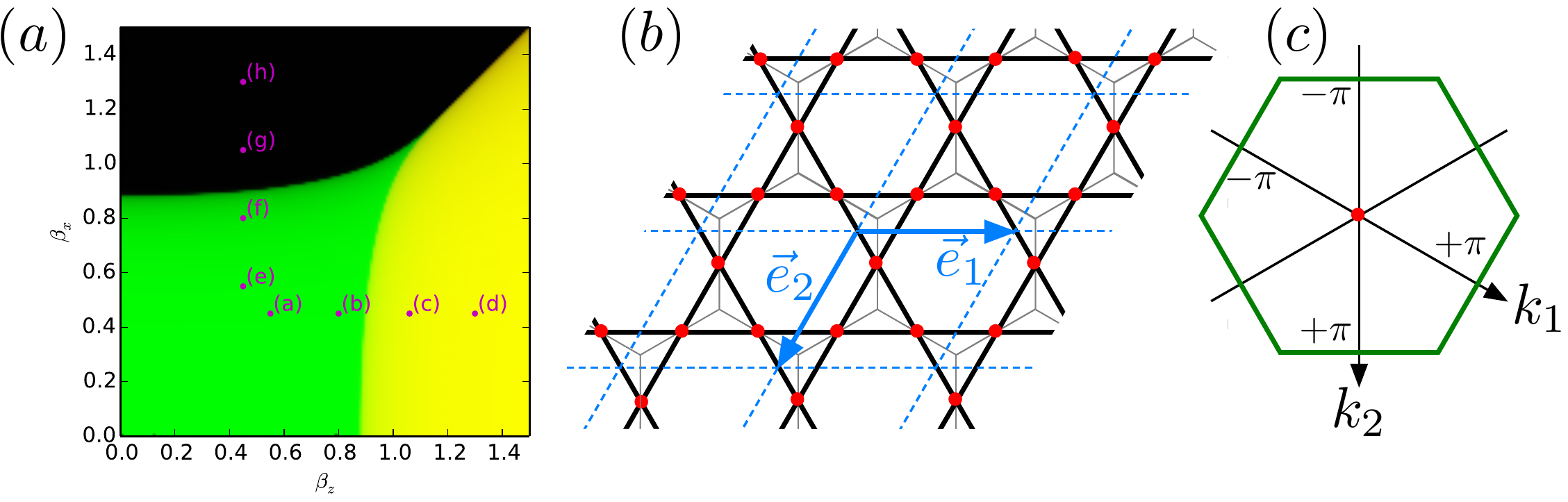}
\caption{\label{fig:combined}
(a) Phase diagram of the filtered toric code state, where the expectation value of the virtual order parameter $X\otimes X$ is used as green intensity and of $X\otimes \openone$ as red intensity. (Note that red and green combines to yellow in RGB images). (b) Kagome lattice (black) for the spins (red dots) of the RVB state, with indication of the blocking (blue lines) used in the PEPS ansatz, with lattice vectors $\vec{e}_1$ and $\vec{e}_2$ (blue arrows). (c) Associated Brillouin zone with indication of the corresponding momenta $k_1$ and $k_2$.}
\end{figure}

At the virtual level of the PEPS, we can use local order parameters $X\otimes \openone$ and $X\otimes X$ to detect the symmetry breaking of $Z\otimes Z$ and $Z\otimes \openone$ respectively in the fixed point subspace, where $X$ is an operator such that $XZ=-ZX$. Fig.~\ref{fig:combined}(a) shows the structure of the fixed point subspace. The topological `toric code' phase is characterized by a doubly degenerate fixed point which is invariant under the subgroup $(\openone\otimes \openone,Z^{\otimes N_x}\otimes Z^{\otimes N_x})$ but breaks the symmetry under the action of $Z^{\otimes N_x} \otimes \openone$. Creating a physical excitation with flux quantum number 0 (string of identities) or 1 (string of $Z$'s) is manifested at the virtual level as a boundary state in the topologically trivial or, respectively, topologically non-trivial sector. That is, excitations with non-trivial flux correspond to domain wall excitations at the level of the transfer matrix. The charge of the physical excitation can be measured as a charge difference between the ket and bra level of the boundary state using the $Z^{\otimes N_x}\otimes Z^{\otimes N_x}$ operator (which is a symmetry, i.e. eigenvalue 1, for the fixed point subspace).

\begin{figure}
\includegraphics[width=\columnwidth]{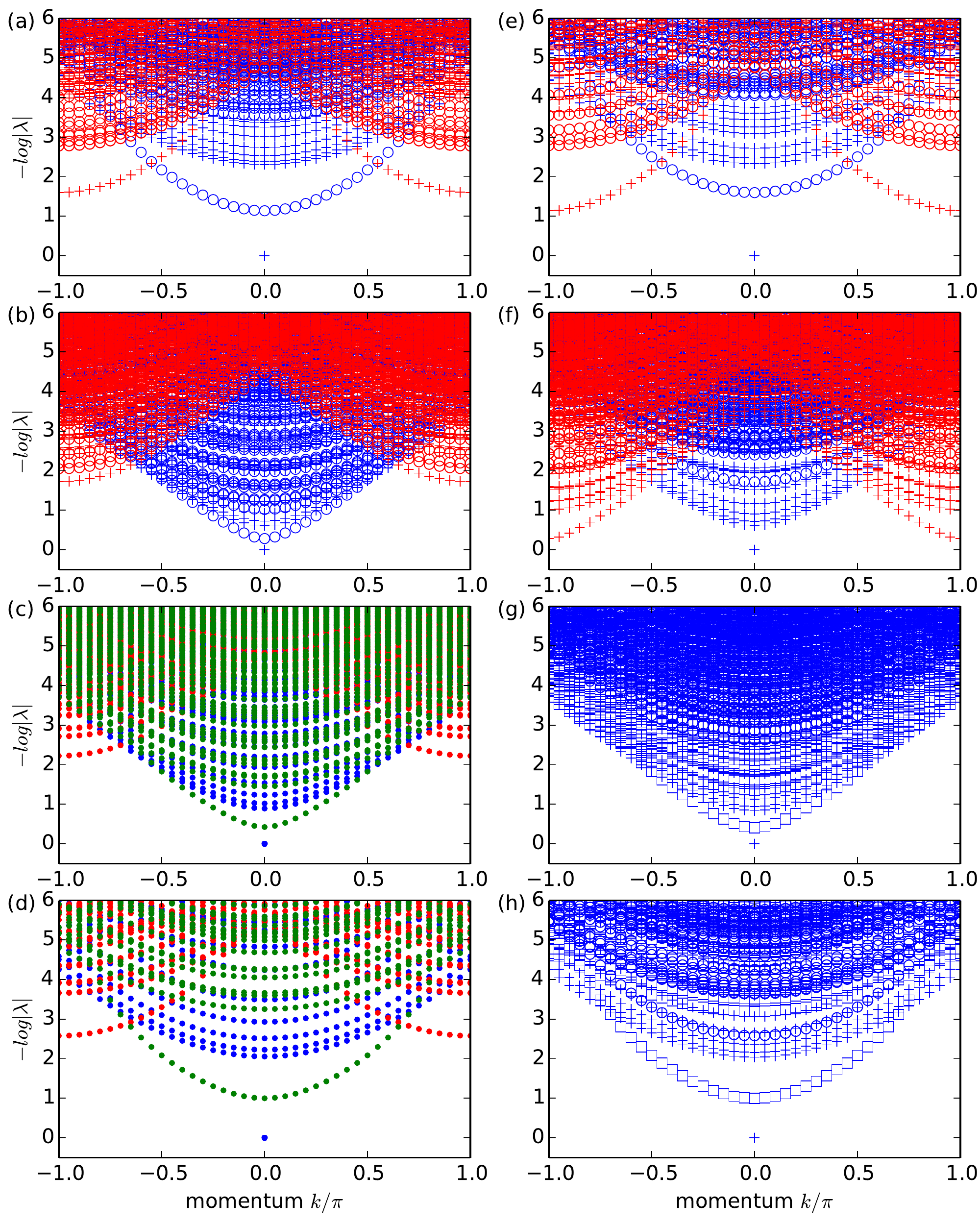}
\caption{\label{fig:spectra}
Spectra of (minus logarithm of) the eigenvalues $\lambda$ of the transfer matrix as function of momentum $k$ for the different points $(\beta_z,\beta_x)$ indicated by the markers (a) - (f) in Fig.~\ref{fig:combined}(a). In the topological phase (a,b,e,f), colors indicate topologically trivial (blue) or non-trivial (red) excitations (flux difference between ket and bra), while the symbol refers to equal (plus sign) or unequal (circle) charge between ket and bra level. In the charge condensed phase (c,d), charges can no longer be measured (dot symbols) and the $Z^{\otimes N}\otimes Z^{\otimes N}$ symmetry is broken. This results in a new topologically non-trivial excitation (green) with a string in both ket and bra and indicates that flux excitations become confined, because there is a nonzero string tension. In the flux condensed phase, the full symmetry is restored and no domain wall excitations of the transfer matrix exist, since they are equivalent to local excitations. In addition, charge can be measured in both ket and bra separately. The states with charge difference (circle) can have individual ket and bra charges $+,-$ and $-,+$ but remain degenerate. States with no charge difference can have ket and bra charges $+,+$ (plus sign) or $-,-$ (square) and the higher energy of the latter indicates string tension between charges and thus charge confinement.}
\end{figure}

Following Ref.~\onlinecite{2014arXiv1408.5140Z}, we can now probe the dispersion relation of the elementary excitations of the model by inspecting the spectrum of (minus logarithm of) the eigenvalues of the transfer matrix in the different sectors, for which we use the one-dimensional excitation ansatz of Ref.~\onlinecite{2012PhRvB..85j0408H}. The result is illustrated for various values of $\beta_z$ and $\beta_x$ inside and outside the topological phase in Fig.~\ref{fig:spectra}. In the topological phase [plots (a), (b), (e) and (f)], the eigenvectors can be labelled by the charge difference between ket and bra (corresponding to the physical charge) and the absence or presence of a half-infinite string (corresponding to the physical flux). We can then relate the spectrum of the transfer matrix to the dispersion relation within the four physical topological sectors and recognize the charge and flux as elementary excitations (isolated branch).  In the first column of Fig.~\ref{fig:spectra}, the gap in the charge sector closes, resulting in charge condensation and a phase transition to the trivial phase [plots (c) and (d)]. Here, the fixed point subspace breaks the full symmetry $(\openone\otimes \openone, Z^{\otimes N_x} \otimes \openone, \openone\otimes Z^{\otimes N_x}, Z^{\otimes N_x}\otimes Z^{\otimes N_x})$ of the transfer matrix and the charge differences between ket and bra is no longer well defined. Correspondingly, we can create new topologically non-trivial excitations with a half-infinite string of $Z$'s in both ket and bra. That this sector has a gap $\Delta_{ZZ}$ indicates that physical flux excitations can no longer exist in isolation and must be confined to pairs, since their normalization goes down as $\exp(-\Delta_{ZZ} L)$ with $L$ the length of the string. In the right column of Fig~\ref{fig:spectra}, the gap in the flux sector closes, corresponding to physical flux condensation. This triggers a phase transition to the trivial phase and results at the virtual level in a unique fixed point with the full symmetry of the transfer matrix. Correspondingly, there are no more topologically non-trivial excitations and we can now measure individual charge numbers of the ket and the bra level. That there is a gap in the $(-,-)$ sector with negative charge in ket and bra indicates that physical charge excitations can no longer exist in isolation and must be confined. 


\subsection{Anyon excitations in the resonating valence bond state on the hexagonal lattice}
\begin{figure}
\includegraphics[width=\columnwidth]{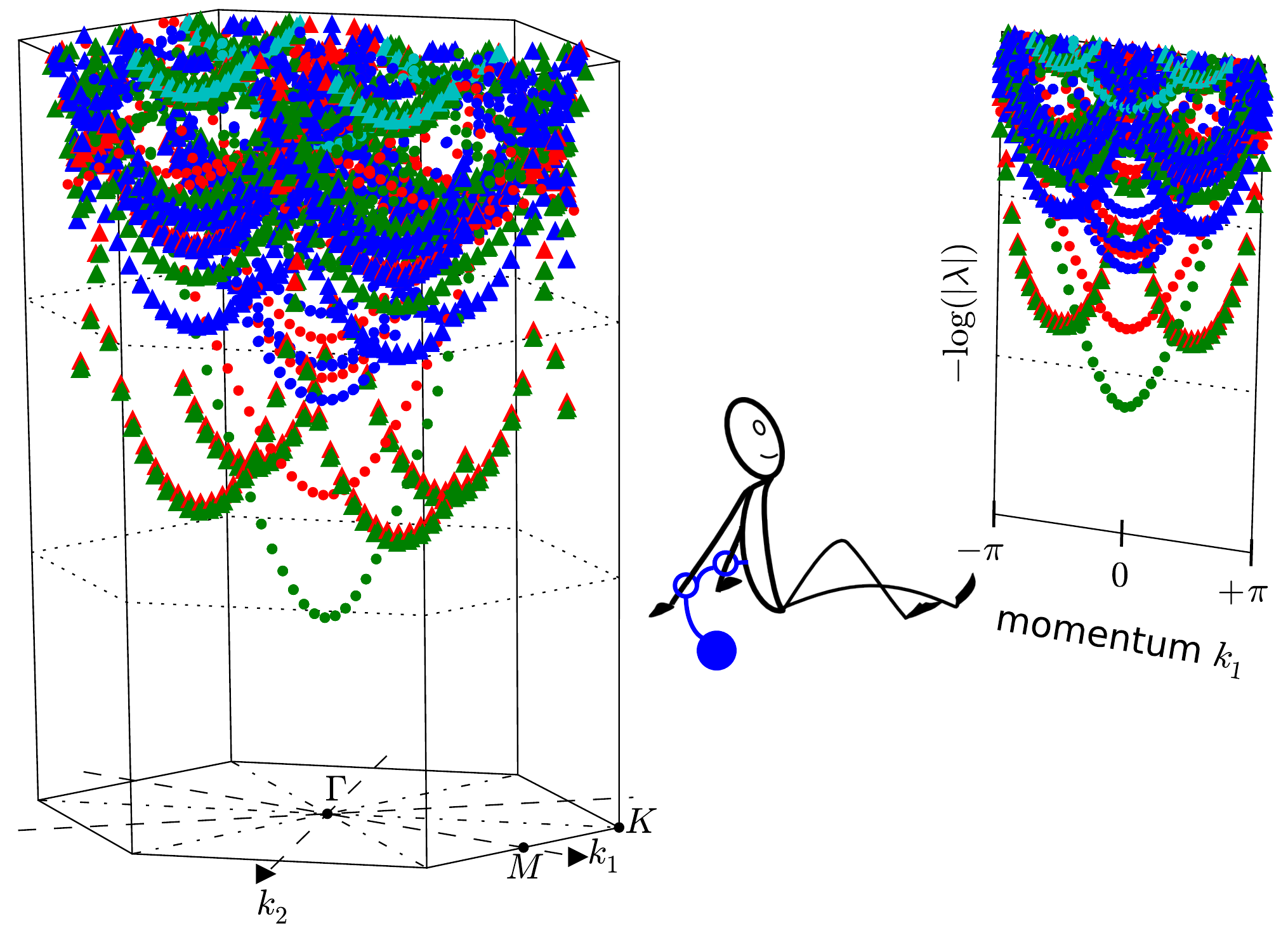}
\caption{\label{fig:rvbspectrum}
A condensed matter physicist in Plato's cave, chained by tensor networks, is able to extract the anyons in the real world from their shadow in the spectrum of the transfer matrix. Colors indicate total spin $S$ (red: $0$, green: $1/2$, blue: $1$, cyan: $3/2$) whereas markers indicates eigenvalues in the trivial (dot) or non-trivial (triangle) sector.}
\end{figure}

Finally, in Fig.~\ref{fig:rvbspectrum} we present the spectrum of the transfer matrix for the resonating valence bond (RVB) state \cite{Anderson:1987aa} on the Kagome lattice, for which the PEPS is also $\mathbb{Z}_2$-injective \cite{PhysRevB.86.115108,PhysRevB.87.140407}. The Kagome lattice was blocked as illustrated in Fig.~\ref{fig:combined}(b), and the eigenvalues of the transfer matrix along the lattice vector $\vec{e}_1$ were computed, giving full access to the momentum $k_1$. The phase of these eigenvalues can then be interpreted as momentum $k_2$, which allows to map them to the Brillouin zone according to Fig.~\ref{fig:combined}(c). This clearly allows to extract the physical elementary excitations \cite{PhysRevB.40.7133,PhysRevB.39.259}, namely the spinons ($S=1/2$, no string) and visons ($S=0$, string), but also vison-spinon bound states ($S=1/2$, string) which seem to occur at energies lower than the vison energies. Finally, there are also an isolated branch of trivial spinless excitations ($S=0$, no string) which could be a bound state of either a vison pair or a spinon pair. For this model, the transfer matrix is a non-hermitian matrix product operator and one can question the validity of an excitation ansatz based on a local perturbation on top of the fixed point (which is approximated as MPS), as this ansatz is only provably justified for the case of local Hamiltonians \cite{PhysRevLett.111.080401}. Numerical evidence for the validity of our results is provided in the Supplementary Material.

\section{Discussion}

We have illustrated how the eigenvalue spectrum of the one-dimensional (quantum) transfer matrix provides a holographic description of the dispersion relations of elementary excitations in the full two-dimensional quantum system. This holds true even in systems with topological order, where the elementary excitations are anyons. The presense of topological order gives rise to particular (virtual) symmetries of the transfer matrix. By carefully studying the manifestation of these symmetries in the fixed point subspace, we were able to relate the different topological sectors of the physical excitations to corresponding topologically non-trivial symmetry sectors (domain walls) at the virtual level. This shows, in particular, that the existence of anyon excitations requires a particular type of symmetry breaking of the doubled virtual symmetry in the fixed point subspace of the transfer matrix, whereas topological phase transitions give rise to a fixed point subspace with a larger or smaller degeneracy. 

While these results might be reminiscent of the closely related bulk-edge correspondence observed in chiral topological phases \cite{PhysRevB.41.12838,Wen:1994aa}, we would like point out the subtle differences. In the PEPS formalism, the properties of the edge states are determined by the fixed points of the transfer matrix \cite{PhysRevLett.112.036402}, whereas here we explicitly consider the complete (long-distance) spectrum of the transfer matrix. In addition, the framework for characterizing topological order in PEPS using MPOs, which is of central importance for our results, has so far only been made explicit for the non-chiral string net models, and it remains to be clarified how the recently discovered chiral PEPS \cite{2013arXiv1307.7726D,PhysRevLett.111.236805,2015PhRvL.114j6803Y} fit within this framework.

This technique holds a powerful potential for studying fundamental questions of topological order and topological phase transitions. While we have studied transfer matrices originating from a tensor network representation of the ground state, the results presented in this paper should generalize to the full quantum transfer matrix obtained from representing the ground state as an imaginary time path integral. Whereas the exact path integral representation can have a gauge theory as virtual boundary, the PEPS truncation will eliminate the gauge degrees of freedom. Correspondingly, the local order parameter measuring the symmetry breaking transitions at the PEPS virtual level will map to a string operator in the temporal direction of the full path integral.

\begin{acknowledgements}
This work was initiated during the program on `Quantum Hamiltonian Complexity' held at the Simons Institute for the Theory of Computing. We acknowledge discussions with Nick Bultinck, Ignacio Cirac, Micha\"{e}l Mari\"{e}n, Burak Sahinoglu, Steve Simon, Joost Slingerland, Laurens Vanderstraeten, Karel Van Acoleyen and Dominic Williamson.
We gratefully acknowledge support by EU grants SIQS and QUERG, the Austrian FWF SFB grants FoQuS and ViCoM. We further acknowledge the support from the Research Foundation Flanders (J.H., F.V.) and the Alexander von Humboldt foundation (N.S.), and computational resources provided by JARA-HPC via grants jara0084 and jara0092.
\end{acknowledgements}

\appendix

\clearpage

\section*{Supplementary Material}
In this supplementary material we reconsider the transfer matrix spectrum for the resonating valence bond state on the Kagome lattice. We elaborate on the construction of the mixed transfer matrix on the cylinder and present the corresponding spectra obtained using exact diagonalization. Then, we revisit the question regarding the validity of the excitation ansatz used in the main text, by comparing the data on the infinite plane obtained using the ansatz with the cylinder data obtained using exact diagonalization, and by providing another measure to test the accuracy of the eigenvector approximation.

First, however, we will discuss explicitly how to construct the mixed transfer operator in the case of 1D systems. A Matrix Product State $\ket\psi = \sum_{s_1,\dots,s_L} \mathrm{tr}[A^{s_1}\cdots A^{s_N}]$ with topologically non-trivial excitations, this is, symmetry breaking, is described by block-diagonal tensors $A^s = \oplus_i A^{s,i}$.  The symmetry broken ground states $\ket{\psi_i}$ are then described by the individual $A^{s,i}$, and domain wall excitations are obtained by switching between different $i$ in the matrix product (alternatively, this can be achieved by attaching a string operator which transforms a product of $A^{s,i}$ to one of $A^{s,i'}$).  In the overlap of the momentum superpositions of these domain walls, the mixed transfer operator $\mathbb{T}^i_{i'} = \sum_{s} A^{s,i}\otimes \bar A^{s,i'}$ appears between the ket and bra position of a domain wall between sectors $i$ and $i'$. Its leading eigenvalue $\lambda$ thus characterizes the correlation function of a pair of such excitations and can be used to infer the momentum $\arg(\lambda)$ and the quasi-energy $-\log|\lambda|$ of the minima of the dispersion.

One can additionally assign quantum numbers of non-broken symmetries to these excitations.  To this end, consider a symmetry $\ket\psi = U_g^{\otimes L} \ket\psi$ (for simplicity, we restrict to finite groups $\mathsf{G}\ni g$). Then, it holds that $\sum_t (U_g)_{st} A^{t} = V_g^\dagger A^sV_g$~\cite{2008PhRvL.100p7202P}, where $V_g$ inherits the algebraic structure of $U_g$ (more precisely, it is an induced projective representation~\cite{2011PhRvB..84p5139S}). The quantum number of an excitation is then given by the quantum number of the corresponding left eigenvector of the (mixed) transfer operator with respect to\ $V_g\otimes \bar{V}_g$.

\begin{figure}
\includegraphics[width=0.9\columnwidth]{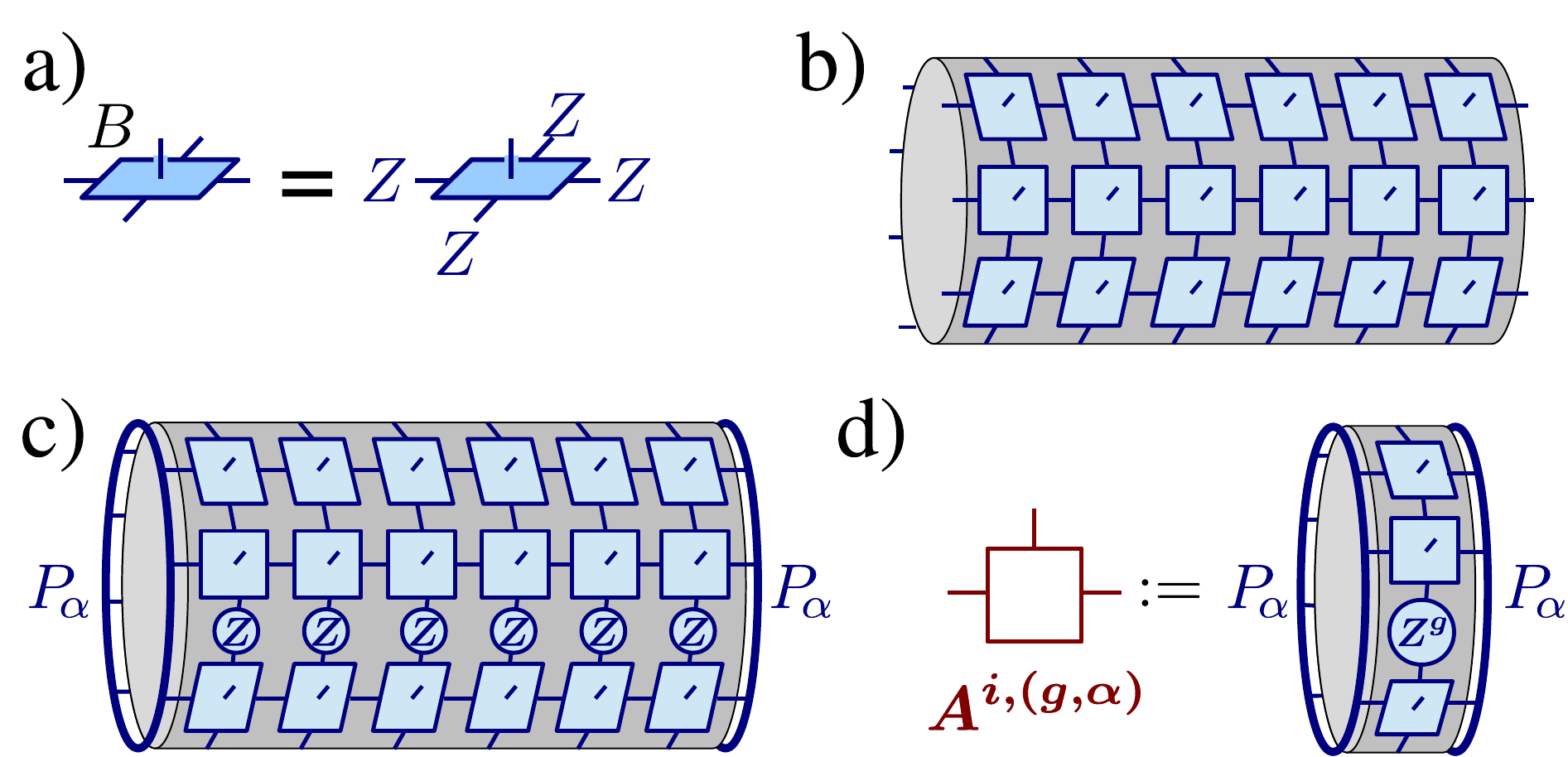}
\caption{
\label{fig:z2-tensors}
(a) Symmetry property of a $\mathbb{Z}_2$ injective PEPS tensor (b) PEPS on the infinite cylinder; (c) Construction of different topological sectors; (d) Construction of the corresponding MPS tensor.
}
\end{figure}

Let us now turn towards 2D cylindrical systems. Both the (perturbed) toric code and the resonating valence bond states discussed in the main text have a $\mathbb{Z}_{2}$-injective PEPS representation. The tensors then have a $\mathbb Z_2$ symmetry on the auxiliary indices, as illustrated in Fig.~\ref{fig:z2-tensors}(a). We consider states on an infinite cylinder with circumference $N_x$; the overall state is then obtained by arranging the tensors on the cylinder and contracting the auxiliary indices as in Fig.~\ref{fig:z2-tensors}(b). We denote the tensor obtained by blocking all tensors in one column (which yields an MPS) by $A$. The four minimally entangled topological sectors, corresponding to the four different quasi-particles, are obtained by\emph{(i)} placing a string of $Z^g$ ($g=0,1$) along the cylinder axis and \emph{(ii)} choosing boundary conditions supported in the irreducible representations $\alpha=\pm 1$ of $Z^{\otimes N_x}$, as depicted in Fig.~\ref{fig:z2-tensors}(c). By interpreting this system as a one-dimensional chain, we can then use the symmetry of the PEPS tensor to construct four independent blocks of the MPS tensor $A^{i,(g,\alpha)}$ by blocking a column as in Fig.~\ref{fig:z2-tensors}(d). The corresponding mixed transfer operator is given by
\[
{\mathbb T}_{g,\alpha}^{g',\alpha'}=\sum_s A^{s,(g,\alpha)}\otimes \bar
A^{s,(g',\alpha')}\ .
\]
As explained in the main text (see Fig.~\ref{fig:alternative}), the value of $g$ and $g'$ determines the presence of a string in ket and or bra of the corresponding domain wall excitations of the normal transfer matrix. However, in the thermodynamic limit of the topological phase, the symmetry of the fixed point under $Z^{\otimes N_x}\otimes Z^{\otimes N_x}$ ensures that the results for $(g,g')\to(g+1\mod 2,g'+1\mod 2)$ are exactly degenerate, so that the results only depend on $g-g'\mod 2$. In addition, because the fixed point subspace of the transfer matrix breaks the $Z^{\otimes N_x}\otimes \openone$ symmetry, only the difference $\alpha-\alpha' \mod 2$ can be measured. While it is possible to build fixed points with well defined quantum numbers for $\alpha$ and $\alpha'$ individually, the pairs $(\alpha,\alpha')$ and $(\alpha+1\mod 2,\alpha'+1\mod 2)$ will also be exactly degenerate, and only depend on the difference $\alpha-\alpha' \mod 2$. Note that on a cylinder with finite circumference $N_x$, the degeneracy will only manifest itself up to exponentially small corrections in $N_x$.

\begin{figure}
\includegraphics[width=0.38\columnwidth]{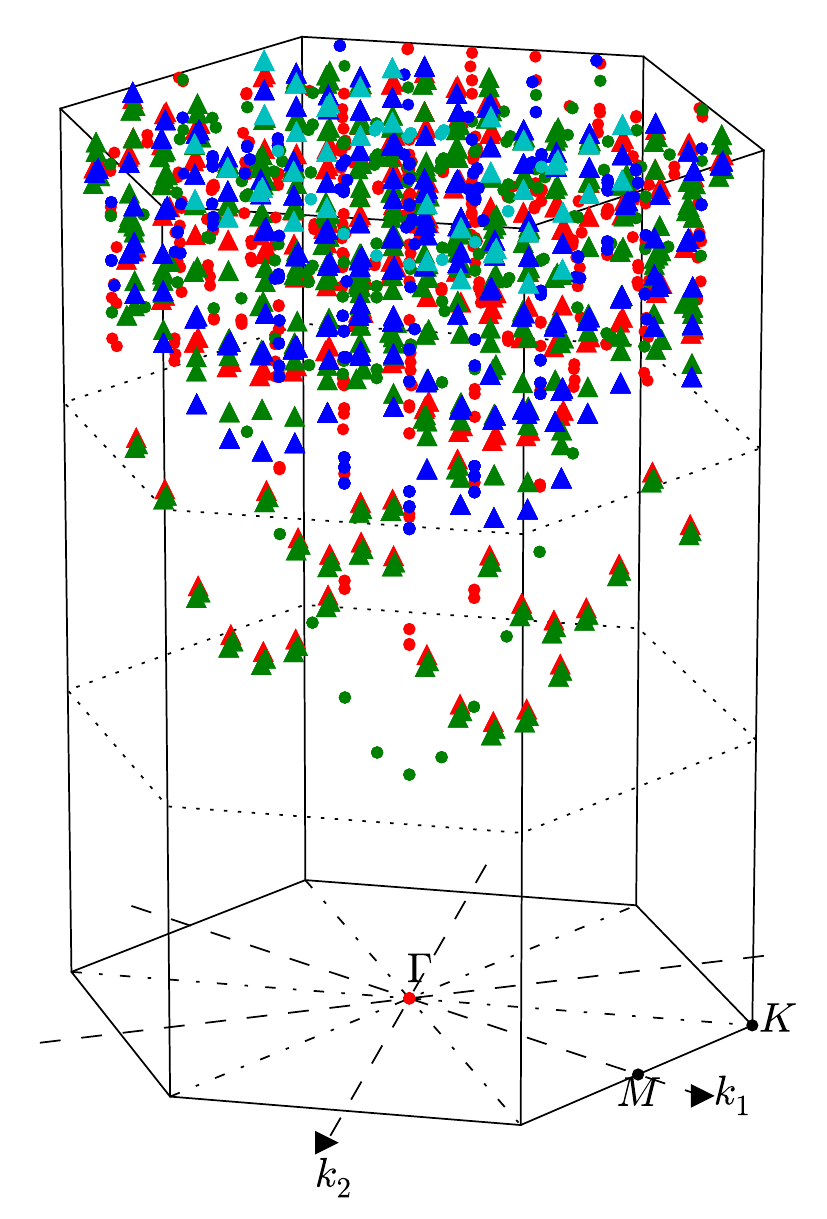}
\includegraphics[width=0.58\columnwidth]{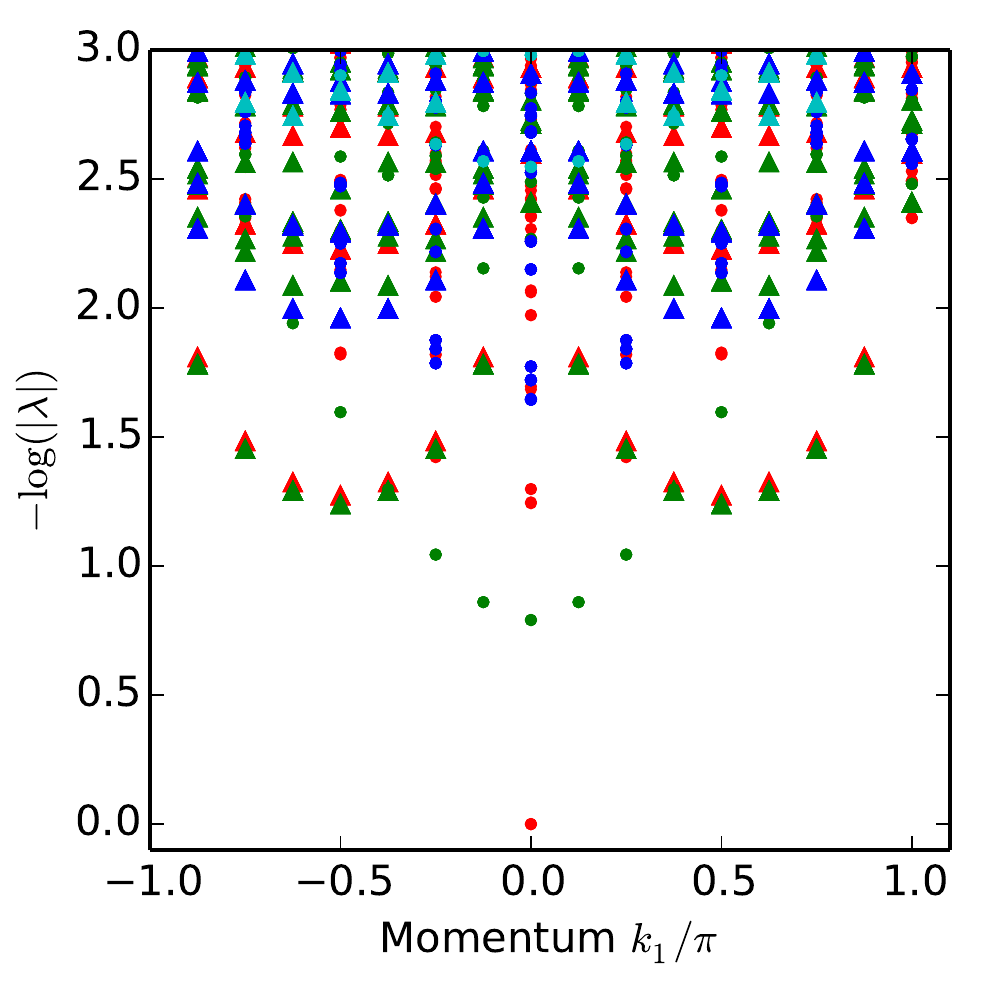}
\caption{\label{fig:rvbspectrumcylinder}
Spectrum of the RVB transfer matrix obtained on a cylinder with circumference $N_x=8$. Markers indicate $g-g'\mod 2 = 0$ (dots) or $g-g'\mod 2 = 1$ (triangles), whereas colors indicate the spin $S$ of the excitation (red: $S=0$, green: $S=1/2$, blue: $S=1$, cyan: $S=3/2$). The value of $\alpha-\alpha'$ is superseded by the spin quantum number, since all integer spin excitations have $\alpha-\alpha'\mod 2= 0$ and all half integer spin excitations have $\alpha - \alpha' \mod 2 = 1$.}
\end{figure}

By blocking the cylindrical PEPS into an MPS, the translation invariance in the $x$ direction (transversal direction) becomes an on-site symmetry, and we can still label the eigenvectors of the transfer matrix with the corresponding quantum numbers. To this end, we need to determine the virtual action $\tau$ of the physical translation operator $T_x$; the momenta are then given by the eigenvalues of $\tau\otimes \bar \tau$ evaluated on the corresponding eigenvector of $\mathbb T_{g,\alpha}^{g',\alpha'}$. This can be done by considering Fig.~\ref{fig:virtualtranslation} for this particular case. As one can see, the action of $\tau$ corresponds to a translation of the virtual indices. However, if a string of $Z$'s is present (i.e. $g=1$), this string is moved as well, and restoring it at its original position requires an additional multiplication to be performed by $Z$, i.e., $\tau = T\cdot Z_1$, where $T$ is the regular translation operator and $Z_1$ acts on the spin next to the location of the string. This leads to an important consequence for the labeling of momenta. We have that $\tau^{N_x} = (Z^g)^{\otimes N_x}$, which implies that $\tau^{N_x}$ has eigenvalue $-1$ when both $g=1$ and $\alpha=-1$. Correspondingly, $(\tau\otimes \bar \tau)^{N_x} = Z^{\otimes N_x}\otimes Z^{\otimes N_x} = \pm 1$, with eigenvalue $-1$ corresponding to the sectors $(g=1\neq g',\alpha=1)$, $(g\neq 1=g',\alpha'=1)$ or $(g=g'=1,\alpha\neq \alpha')$. In those sector, the eigenvalues $e^{ik}$ of $\tau\otimes \bar \tau$ are shifted by half a spacing, i.e., $k_x=2\pi(n+\tfrac12)/N_x$, with $n=0,\dots,N_x-1$. This can be checked for the data presented in Fig.~\ref{fig:rvbspectrumcylinder} for $N_x=8$. The fact that this results in smooth dispersion relations for $2N_x$ instead of $N_x$ different momenta is truly remarkable.

\begin{figure}
\includegraphics[width=0.6\columnwidth]{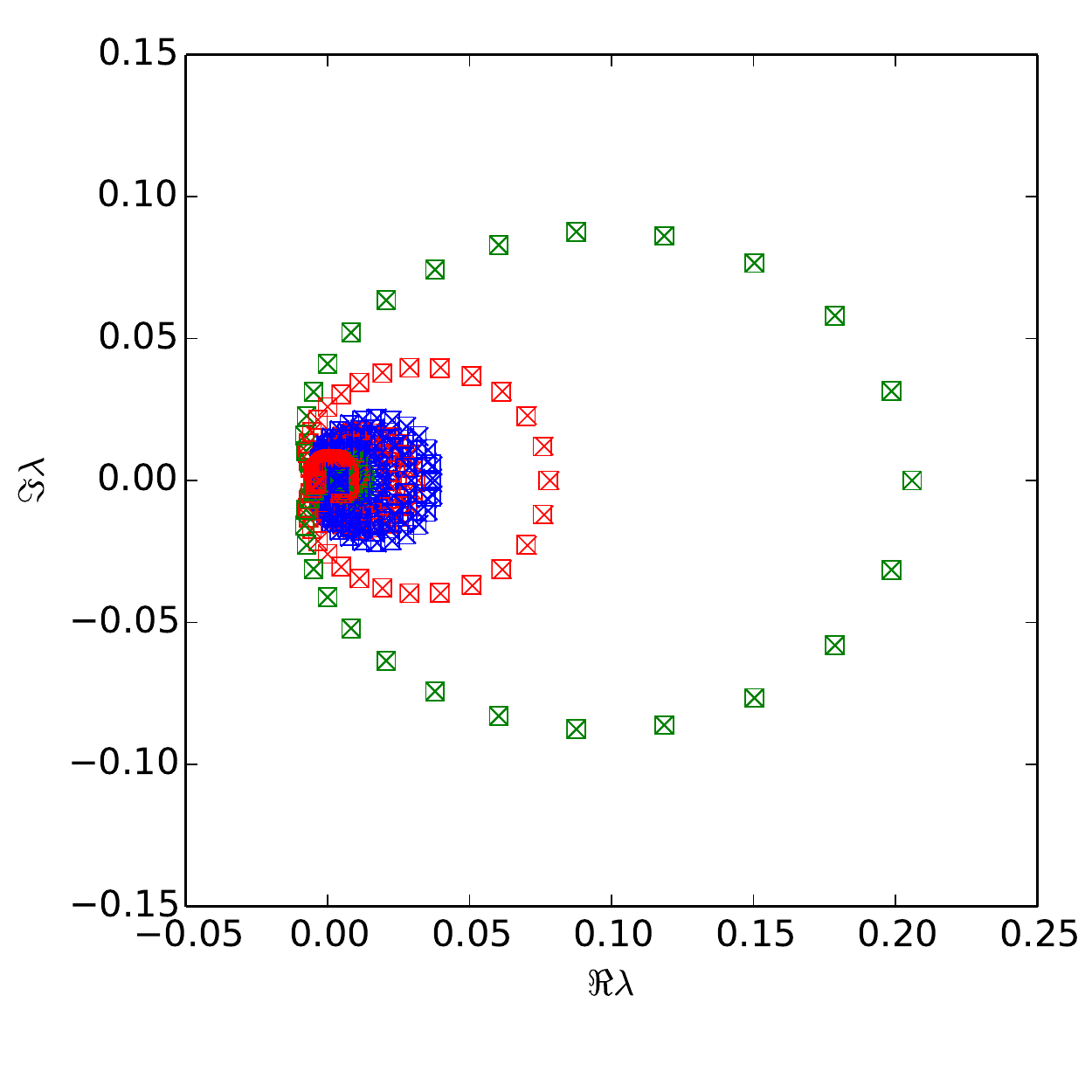}
\caption{\label{fig:comparison}
Comparison of the eigenvalues of $(P\mathbb{T} P)^2$ (squares) with the eigenvalues of $P \mathbb{T}^2 P$ (crosses), where $P$ is the projector onto the subspace spanned by the variational excitation ansatz in the trivial sector. The fact that these eigenvalues collapse for the eigenvalues of largest magnitude confirms that the variational subspace can accurately capture the exact eigenvectors of $\mathbb{T}$ corresponding to the largest eigenvalues (lowest eigenvalues in Fig.~\ref{fig:rvbspectrum}).}
\end{figure}

Finally, we can compare the exact diagonalization results of Fig.~\ref{fig:rvbspectrumcylinder} with the results obtained using the variational excitation ansatz in the limit $N_x\to\infty$, as presented in Fig.~\ref{fig:rvbspectrum} of the main text. These results match perfectly, where of course the results in the main text do not suffer from finite size effects (but perhaps from finite $D$ effects in the MPS approximation) and can have an arbitrary resolution in momentum space, since all quantum numbers $k_1\in[-\pi,+\pi]$ are available in the thermodynamic limit. The variational excitation ansatz defines, for every momentum slice, a linear subspace of the Hilbert space. If $P_k$ denotes the orthogonal projector onto this subspace, we are essentially computing the eigenvalues of $P_k \mathbb{T} P_k$ instead of the eigenvalues of $\mathbb{T}$, which could be very different if the variational subspace would not be able to accurately capture exact eigenvectors of $\mathbb{T}$. As a further confirmation, we compare in Fig.~\ref{fig:comparison} the eigenvalues of $(P\mathbb{T} P)^2$ with the eigenvalues of $P \mathbb{T}^2 P$, where $P$ is the projector onto the total excitation subspace $P=\int \mathrm{d}k P_{k}$. 

\clearpage

\end{document}